# More chips off of Asteroid (4) Vesta: characterization of eight Vestoids and their HED meteorite analogs

Authors: Paul S. Hardersen [a,*,1], Vishnu Reddy [b,*], Rachel Roberts [c,*], Amy Mainzer [d]

[a] University of North Dakota, Department of Space Studies, 4149 University Avenue, Stop 9008, 530 Clifford Hall, Grand Forks, North Dakota, USA 58202-9008, Hardersen@space.edu.

[b] Planetary Science Institute, 1700 E. Fort Lowell Road, Suite 106, Tucson, Arizona, USA 85719, reddy@psi.edu.

[c] University of North Dakota, Department of Space Studies, 4149 University Avenue, Stop 9008, 521 Clifford Hall, Grand Forks, North Dakota, USA 58202-9008, momenttensor@gmail.com.

[d] Jet Propulsion Laboratory, 4800 Oak Grove Drive, Pasadena, California, USA 91109, amy.mainzer@jpl.nasa.gov.

[*] Visiting Astronomer at the Infrared Telescope Facility, which is operated by the University of Hawai'i under Cooperative Agreement No. NNX-08AE38A with the National Aeronautics and Space Administration, Science Mission Directorate, Planetary Astronomy Program.

[1] Corresponding author. *Email address:* Hardersen@space.edu (P.S. Hardersen).

# I. Abstract

Vestoids are generally considered to be fragments from Asteroid (4) Vesta that were ejected by past collisions that document Vesta's collisional history. Dynamical Vestoids are defined by their spatial proximity with Vesta (Zappala et al., 1995; Nesvorny et al., 2012). Taxonomic Vestoids are defined as V-type asteroids that have a photometric, visible-wavelength spectral, or other observational relationship with Vesta (Tholen, 1984; Bus and Binzel, 2002; Carvano et al., 2010). We define 'genetic Vestoids' as V-type asteroids that are probable fragments ejected from (4) Vesta based on the supporting combination of dynamical, near-infrared (NIR) spectral, and taxonomic evidence. NIR reflectance spectroscopy is one of the primary ground-based techniques to constrain an asteroid's major surface mineralogy (Burns, 1993a). Despite the reasonable likelihood that many dynamical and taxonomic Vestoids likely originate from Vesta, ambiguity exists concerning the fraction of these populations that are from Vesta as compared to the fraction of asteroids that might not be related to Vesta.

Currently, one of the most robust techniques to identify the genetic Vestoid population is through NIR reflectance spectroscopy from ~0.7-2.5 μm. The derivation of spectral band parameters, and the comparison of those band parameters with those from representative samples from the Howardite-Eucrite-Diogenite (HED) meteorite types, allows a direct comparison of their primary mineralogies. Establishing tighter constraints on the genetic Vestoid population will better inform mass estimates for the current population of probable Vestoids, will provide more accurate orbital information of Vestoid migration through time that will assist dynamical models, and will constrain the overall current abundance of basaltic material in the main asteroid belt (Moskovitz et al., 2008).

This work reports high-quality NIR spectra, and their respective interpretations, for eight $V_p$-type asteroids, as defined by Carvano et al. (2010), that were observed at the NASA Infrared Telescope Facility on January 14, 2013 UT. They include: (3867) Shiretoko, (5235) Jean-Loup, (5560) Amytis, (6331) 1992 FZ1, (6976) Kanatsu, (17469) 1991 BT, (29796) 1999 CW77, and (30872) 1992 EM17. All eight asteroids exhibit the broad ~0.9- and ~1.9-μm mineral absorption features indicative of pyroxene on each asteroid's surface. Data reduction and analysis via multiple techniques produced consistent results for the derived spectral absorption band centers and average pyroxene surface chemistries for all eight asteroids (Reddy et al. 2012c; Lindsay et al., 2013, 2014; Gaffey et al., 2002; Burbine et al., 2009). (3867) Shiretoko is most consistent with the eucrite meteorites while the remaining seven asteroids are most consistent with the howardite meteorites. The existing evidence suggests that all eight of these $V_p$-type asteroids are genetic Vestoids that probably originated from Vesta's surface.

## 1. Introduction

### *1.1. Vesta, Vestoids, and V-type asteroids*

Much of the past work on V-type asteroids has centered on (4) Vesta because this large, main-belt asteroid was recognized early on as a unique object based on its size and its early linkage to the HED meteorites (McCord et al., 1970). Vesta was also considered unusual in the early asteroid taxonomies (Chapman et al., 1975; Bowell et al., 1978). Subsequent workers extended the taxonomic classification to surface mineralogical studies. McCord et al. (1970) made the first direct connection between (4) Vesta and the HED meteorites while Gaffey (1983, 1997) characterized Vesta's surface as predominantly howarditic from analysis of a ground-based NIR reflectance spectrum. In that later work, Gaffey (1997) identified some surface mineralogic variations across Vesta's surface that were correlated with possible large impact sites.

Renewed focus on (4) Vesta in the 1990s occurred after the discovery of a large, south pole impact basin from Hubble Space Telescope imagery (Binzel et al., 1997; Thomas et al., 1997). This discovery provided greater insight into Vesta's collisional history, introduced the possibility of probing deeper crustal units beneath the surface that were excavated by the impact(s), and gave supporting evidence for the prior discovery of a Vesta-like population of small basaltic asteroids stretching from Vesta's orbit ($a \sim 2.36$ AU) to the 3:1 mean-motion resonance (~2.50 AU) (Binzel and Xu, 1993). Comparison of rotationally resolved NIR spectra of Vesta from the NASA IRTF suggested that albedo variations might be linked to surface compositional heterogeneity (e.g., Gaffey, 1997; Vernazza et al., 2005; Reddy et al., 2010). Later Hubble Space Telescope (HST) observations confirmed the affinity between compositional variations and albedo units thought to be surface morphological features (Thomas et al., 1997; Binzel et al., 1997; Li et al., 2010).

Recent results from NASA's Dawn mission at Vesta provide high-resolution visual and spectroscopic evidence of the asteroid's surface mineralogy and collisional history (De Sanctis et al., 2012; Marchi et al., 2012; Reddy et al., 2012a,b; Russell et al., 2012; Schenk et al., 2012; McSween et al., 2013). Schenk et al. (2012) described two large south pole impact basins, Veneneia and Rheasilvia, which record different events with Veneneia recording Vesta's first south pole impact. From a global perspective, Marchi et al. (2012) identified 1872 craters with diameters larger than 4 km that are distributed across Vesta's surface. Among this population were 12 craters with diameters greater than 50 km scattered across Vesta's equatorial belt (Marchi et al., 2012), which is the cratering population most likely to produce ejecta (i.e., Vestoids) in the inner solar system.

Vesta's importance derives from its provenance, differentiation, collisional evolution, and status as one of the few remaining intact planetesimals since solar system formation. This distinction is reinforced by Vesta's igneous nature, which is unique when compared to the compositional and mineralogic (i.e., primitive) constitution of the few other putative, remaining intact main-belt asteroids such as (1) Ceres and (2) Pallas (Drake, 2001; Schmidt and Castillo-Rogez, 2012; Kuppers et al., 2014). Study of (4) Vesta has implications for the nature of the sole early solar system heating event (Herbert et al., 1991; Grimm and McSween, 1993), the collisional evolution of small asteroids via mechanisms such as Yarkovsky and YORP (Bottke et al., 2006), and the question of whether or not Vesta is the only source of basaltic material in the asteroid belt.

Besides Vesta, most studies of V-type asteroids have been associated with the first Vestoids discovered by Binzel and Xu (1993), but subsequent efforts have contributed to the spectral and mineralogical characterizations of a somewhat broader range of potential Vestoids in the visible and/or near-infrared (VNIR) spectral regions (Vilas et al., 2000; Burbine et al., 2001; Kelley et al., 2003; Cochran et al., 2004; Alvarez-Candal et al., 2006; Duffard et al., 2004, 2006; Roig et al., 2008; Duffard and Roig, 2009; Moskovitz et al., 2008, 2010; De Sanctis et al., 2011a,b; Mayne et al., 2011; Reddy et al., 2011a; Solontoi et al., 2012; Jasmim et al., 2013; Hicks et al., 2014).

These previous efforts have largely confirmed that the original Vestoids likely originate from (4) Vesta, but some interesting discrepancies have appeared among the results of various workers. This is the case for (1929) Kollaa, which has been studied by Kelley et al. (2003), Moskovitz et al. (2010), and Mayne et al. (2011). While these works reported consistent Band I absorption centers, the Band II absorption centers are quite divergent and range from 1.914- to 1.980-μm (Kelley et al., 2003; Moskovitz et al., 2010). These band center variations, which could be attributed to surface mineralogical variations as well as other possible causes, produce meteorite analog candidates that range from the diogenites (Kelley et al., 2003) to the eucrites (Moskovitz et al., 2010).

In addition, Reddy et al. (2011b) and De Sanctis et al. (2011b) have identified diogenite-like V-type asteroids. Reddy et al. (2011b) noted that the surface composition of near-Earth asteroid (237442) 1999 TA10 is analogous to near-pure diogenite and suggest it to be a fragment of the lower crust/upper mantle of Vesta.

### *1.2. Research program goals*

The existing knowledge of asteroids, both individually and as discrete collections of related bodies (i.e., the main asteroid belt, Trojans, etc.) follows a general trend that begins with discovery and orbit determination, followed by basic physical characterization (i.e., albedo, diameter, rotation rate, broadband colors, visible-wavelength spectra, etc.), and then to later detailed characterization (i.e., NIR spectra, radar studies, pole determinations, spacecraft visits, etc.). While each asteroid that has been discovered does not follow exactly the same path in terms of the knowledge that is gleaned from it as a function of time, the inflow of information about asteroids follows the general trend from large discovery surveys that lead to later detailed studies of individual bodies. This trend can be seen in the various macroscopic datasets that are available for the minor planets. As of July 12, 2014, the IAU Minor Planet Center reported orbits for 645,148 minor planets, of which 401,810 are numbered minor planets, and 18,702 are named minor planets (IAU Minor Planet Center: http://www.minorplanetcenter.net/iau/lists/ArchiveStatistics.html). Asteroids with rotation rates derived from broadband photometric light curves number 5,954 as of February 28, 2014 (http://www.minorplanet.info/lightcurvedatabase.html).

By comparison, asteroid taxonomic classifications that group asteroids into a variety of optical and spectral groups exist for at least 2,615 asteroids (Neese, 2010). Existing asteroid spectral and radar datasets are even smaller. Visible-wavelength spectral data that are publicly available include ~1500 asteroids (Xu et al., 1995; Vilas et al., 1998; Larson, 2000; Bus and Binzel, 2002; Lazzaro et al., 2004). Radar detections are available for at least ~388 asteroids (Neese et al., 2012). NIR spectra exist for at least ~750 asteroids (e.g., Gaffey et al., 1993; Binzel et al., 2006; Mothe'-Diniz et al., 2008; Reddy, 2010, 2011; Birlan et al., 2011; Bus, 2011; Clark et al., 2004, 2011; De Sanctis et al., 2011a,b; Fornasier et al., 2011; Hardersen et al., 2011; Popescu et al., 2011; Fieber-Beyer, 2013; Thomas et al., 2014). Even though these different datasets continue to grow, the relative dearth of detailed information suggests a need to improve the physical characterizations of most asteroids. This vital information will improve our understanding of both individual asteroids as well as the large populations of small bodies.

*1.2.1. Goal 1 – Confirm the $V_p$-taxonomy – basaltic asteroid linkage*

The initial goal for this research is to test the taxonomic-mineralogic association for a reasonable sample size (~139) of the 650 Wide-Field Infrared Survey Explorer (WISE)/NEOWISE-defined $V_p$-type asteroids (Carvano et al., 2010; Mainzer et al., 2012). The Carvano et al. (2010) taxonomic classes are defined by their low-resolution visible-wavelength reflectance spectra that are derived from Sloan Digital Sky Survey *u'g'r'i'z'* colors. The $V_p$ taxonomic class is characterized by the deepest ~1-μm absorption feature among all of the Carvano et al. (2010) classes, which is the result of a significant reduction in reflectance at the *z'* filter. Mainzer et al. (2012) combined the $V_p$-type asteroids in Hasselmann et al. (2012) with the WISE albedos in Masiero et al. (2011) to produce what we are calling 'WISE-defined $V_p$-type asteroids'. The $V_p$-type asteroids are distinct by exhibiting the highest albedos of all of the Carvano et al. (2010) classes (Mainzer et al., 2012). Constraining our target population based on taxonomic *and* albedo information that is consistent with basaltic asteroid features provides the greatest likelihood of characterizing objects that are basaltic.

The results reported in this paper represent the first part of the overall effort to better characterize the sample population of ~139 $V_p$-type asteroids. While recent workers have stated that an explicit taxonomic-compositional association exists for asteroids (DeMeo and Carry, 2014), previous taxonomic classification efforts avoid claiming this type of linkage (Tholen, 1984). NIR spectral efforts have shown that a wide variety of mineralogic and meteoritic diversity exists within a given taxonomic class, such as (but not only) the S-, M-, and X-type asteroids (Gaffey et al., 1993; Clark et al., 2004; Hardersen et al., 2005, 2011; Fornasier et al. 2010).

Previous visible- and near-infrared (VNIR) spectral work on V-type asteroids has produced a false positive rate of ~9%, where a false positive represents a taxonomically classified V-type asteroid that does not exhibit basaltic surface mineralogic characteristics in its VNIR spectrum (Roig and Gil-Hutton, 2006; Moskovitz et al., 2008). As taxonomic classifications are largely based on non-mineralogic properties, previous work has shown that photometric colors, visible-wavelength spectra, color indices, and other taxonomic classifiers are sometimes not indicative of an asteroid's surface mineralogy (Gaffey et al., 1993; Hardersen et al., 2011). The consequence of this is that an asteroid's taxonomic classification only provides broad constraints for an asteroid's composition that requires additional diagnostic efforts (i.e., NIR spectra) to define surface mineralogy and potential meteorite analogs. The low 9% false positive rate suggests that the V-type taxonomy provides a relatively good method for identifying basaltic asteroids (Roig and Gil-Hutton, 2006; Moskovitz et al., 2008). However, this false positive rate is only valid for a small sample size of 21 asteroids (Roig and Gil-Hutton, 2006; Moskovitz et al., 2008). Obtaining a more robust false positive rate for the V-type main-belt asteroid population will better determine the usefulness of the V-taxonomic class, while also identifying those basaltic asteroids that are directly (i.e., dynamically, spectrally, and mineralogically) associated with (4) Vesta and those V-type asteroids in the outer main belt that are basaltic.

This verification process will occur by obtaining NIR reflectance spectra (~0.7-2.5-µm) for ~124 $V_p$-type asteroids that are dynamically associated with (4) Vesta and ~15 $V_p$-type asteroids that are in the outer main belt with semimajor axes > 2.5 AU (Mainzer et al., 2012). This strategy will help to constrain the genetic Vestoid population in the inner main belt while also potentially identifying basaltic asteroids in the outer main belt. Producing NIR spectra of these asteroids will confirm the presence of the deep ~0.9-µm absorption feature, whose short-wavelength presence appears in both visible-wavelength spectral surveys and in Sloan Digital Sky Survey (SDSS) five broadband color *(u'g'r'i'z')* distributions for $V_p$-type asteroids (Bus and Binzel, 2002; Carvano et al., 2010). NIR spectra of $V_p$-type asteroids will also capture most of the very broad ~1.9-µm feature that will facilitate estimation of the average surface pyroxene chemistry and test linkages to the HED meteorites (Gaffey et al., 2002; Burbine et al., 2009).

*1.2.1.1. Outer main-belt basaltic asteroids*

Knowledge of the abundance, physical nature, mineralogy, petrology, and provenance of outer main belt (*a* > 2.5 AU) basaltic asteroids is currently sparse. (1459) Magnya is currently the only asteroid in the outer main belt confirmed to have a basaltic/HED-like mineralogy (Lazzaro et al., 2000; Hardersen et al., 2004). A few other V-type/basaltic asteroid candidates have been identified from the visible-wavelength spectral work of Roig et al. (2008) and Duffard and Roig (2009). Moskovitz et al. (2008) modeled the distribution of basaltic asteroids throughout the asteroid belt and suggests that the total mass of the outer main-belt basaltic population is similar to dynamical Vestoid mass estimates.

Better constraining the population and mass of outer main-belt basaltic asteroids can potentially provide ‘anchor points’ that inform the radial extent and variability of the early solar system heating event (Herbert et al., 1991; Grimm and McSween, 1993). The leading models (i.e., T Tauri induction heating, $^{26}$Al radiogenic heating) of Herbert et al. (1991) and Grimm and McSween (1993) generally do not predict melting temperatures in the outer main belt, although some variations in the induction models of Herbert et al. (1991) suggest that possibility. Characterizing a non-trivial population of igneous asteroids with semimajor axes > 2.8 AU may prompt a reconsideration of these models and renewed investigation of the early solar system heating event.

### *1.2.2. Goal 2 – Mineralogical characterization of $V_p$-type asteroids*

One of the primary methods for characterizing the surface mineralogies of terrestrial solar system bodies lacking atmospheres is via NIR reflectance spectroscopy. This observational technique, which utilizes the wavelength region from ~0.7-2.5-μm, is the main technique used to remotely study Mercury, Earth’s Moon, and small solar system bodies such as the main-belt asteroids (e.g., Gaffey et al., 1993; Pieters et al., 2000; Burbine et al., 2002; Vernazza et al., 2010; Clark et al., 2011; Emery et al., 2011; Reddy et al., 2011; Lindsay et al., 2013; Rivkin et al., 2013). This technique is useful because it focuses on the spectral region where an object’s primary surface mineralogy usually dominates over secondary effects, such as grain size variations, space weathering, observational variables, etc. (Cloutis et al., 1986; Burns, 1993a). The olivine and pyroxene mineral groups are the most important mafic silicate minerals detectable in this spectral region due to their widespread abundance in solar system materials and the presence of transition metal ions (i.e., $Fe^{2+}$) in their mineral crystal structures, which allows the formation of broad, easily identifiable absorption features (Burns, 1993a,b).

Other minerals are detectable in this spectral region, such as water ice and magnetite, and, to a lesser extent, plagioclase feldspar, phyllosilicates, and other less diagnostic minerals (Rivkin and Emery, 2010; Yang and Jewitt, 2010; Takir and Emery, 2012), but the olivine and pyroxene mineral groups are the dominant minerals found on asteroid surfaces for many taxonomic types (i.e., V, S, M, R, A: Gaffey et al., 1993; Gaffey, 1997; Abell and Gaffey, 2000; Moskovitz et al., 2010; Hardersen et al., 2011; Sanchez et al., 2014).

For the purposes of this work, the primary spectrally active minerals of interest are the pyroxene group of minerals; in particular, these include orthopyroxene, low-Ca clinopyroxene, and high-Ca clinopyroxene (i.e., augites). This is the case because basaltic material, which is considered the primary rock type on V-type asteroid surfaces, is spectrally dominated by pyroxene even though plagioclase feldspar is the other major mineral found in basalts. Plagioclase feldspar has a weak, ephemeral absorption feature at ~1.2-µm that has been interpreted in some V-type asteroid spectra (i.e., Hardersen et al., 2004, 2006).

The HED meteorites, which are considered to be from (4) Vesta (McCord et al., 1970; Gaffey, 1997; Drake, 2001; Russell et al., 2012), have a primary meteoritic and mineralogic progression that systematically varies in abundance from the low-$Fe^{2+}$, high-$Mg^{2+}$ diogenites to the high-$Fe^{2+}$ basaltic eucrites, with the howardites exhibiting an intermediate $Fe^{2+}$-content in the pyroxenes as they are brecciated mixtures of diogenites and eucrites (Gaffey, 1976). The typical geologic scenario invoked is a layering environment with the diogenites formed at crustal depths, and with the cumulate and basaltic eucrites forming at shallow depths and at the surface, respectively (e.g., Mittlefehldt et al., 1998). HED major mineralogy varies by type as orthopyroxene is the only major mineral in diogenites while the basaltic and cumulate eucrites are dominated by low-Ca pyroxene and Ca-rich plagioclase feldspar, and the howardites are polymict breccias that are variable mixtures of eucrites and diogenites (e.g., Mittlefehldt et al., 1998).

Spectral-mineralogical laboratory calibrations enable asteroid-HED meteorite linkages by comparing asteroid and meteorite spectral band parameters (i.e., band centers) and their resulting average surface pyroxene chemistries. The original pyroxene band-band plot by Adams (1974) shows a progressive increase in absorption band centers i.e., Band I and Band II) for orthopyroxenes and (Type B) clinopyroxenes. Calibrations to estimate single pyroxene average chemistries have been developed by Gaffey et al. (2002) and Burbine et al. (2009), which produce relatively robust and consistent results (Hardersen et al., 2011).

Another relevant calibration is the Band I center vs. Band Area Ratio (BAR) plot, which was originally developed by Cloutis et al. (1986) to show how these two spectral variables change with the relative proportion of orthopyroxene and olivine in a two-mineral mixture. This was later modified by Gaffey et al. (1993) in the study of the S-asteroid sub-types, which included a rectangular region in Band I vs. BAR space, at large BAR values, that bounds the parameter space for basaltic achondrites. V-type asteroids that are analogous to the HED meteorites, or basalts from other parent bodies, will plot within this region or at even larger BAR values (Gaffey et al., 1993; Hardersen et al., 2004; this work). Yet another complementary analysis technique that can show HED affinities involves deriving asteroid surface pyroxene chemistries and plotting this data on a pyroxene quadrilateral along with representative data from the HED meteorites (Mittlefehldt et al., 1998).

This work will utilize all of the above techniques, along with complementary dynamical and physical data, which will help to determine if the $V_p$-type asteroids in this work originate from (4) Vesta and have affinities to individual HED meteorite types.

### *1.3. WISE and NEOWISE*

#### *1.3.1. Overview*

The Wide-Field Infrared Survey Explorer (WISE) is a medium-class Explorer mission that operated for two mission phases from January 14, 2010, to February 1, 2011 (Wright et al., 2010; Mainzer et al., 2011). The first phase included the cryogenic portion of the mission, which involved four-band observations at 3.4-, 4.6-, 12-, and 22-μm when the primary arrays and instrumentation were solid-$H_2$ cooled. The primary mission goal was to produce a second-generation, high-resolution infrared sky map, which is an order-of-magnitude improvement compared to the infrared sky map produced by the Infrared Astronomical Satellite (IRAS) in the 1980s (Neugebauer et al., 1984). Besides the all-sky survey, WISE science activities focused on the observations of ultra-luminous infrared galaxies, brown dwarf stars, Active Galactic Nuclei (AGN)/Quasi-Stellar Objects (QSO), and asteroids (Wright et al., 2010).

The second mission phase was called the ‘post-cryogenic phase’ and covered the time period from August 5, 2010, to February 1, 2011, when the secondary cryogenic tank, followed by the primary cryogenic tank, began to exhaust their hydrogen supplies. The consequence of cryogen depletion was the loss of use of the 12- and 22-μm bands due to the warming of the instrument (Masiero et al., 2011). By February 2011, WISE had detected more than 158,000 small bodies throughout the solar system (Mainzer et al., 2011). After placing the spacecraft into hibernation mode, WISE was reactivated in September 2013 and will operate as NEOWISE until early 2017 (Mainzer et al., 2014).

Mainzer et al. (2011) describe NEOWISE, a data-processing enhancement to the WISE data archive facilitating the detection and identification of small bodies through the WISE Moving Object Processing Software (WMOPS). Some of the primary physical parameters derived from NEOWISE infrared detections include estimation of asteroid effective diameters, geometric albedos, and infrared albedos (Masiero et al., 2011). Albedos are derived from application of the Near-Earth Asteroid Tracking Model (NEATM: Harris, 1998) described in Masiero et al. (2011).

*1.3.2. The $V_p$-type asteroid population*

Carvano et al. (2010) developed a new taxonomic classification system for asteroids that is based exclusively on Sloan Digital Sky Survey (SDSS) five-color photometry (*u'g'r'i'z'*) designed to be consistent with the Bus and Binzel (2002) spectral-based taxonomy. The five SDSS broadband colors, which extend from 0.3- to 1.123-μm (Fukugita et al., 1996; Carvano et al., 2010), produce a very low-resolution visible-wavelength spectrum that segregates asteroids into various bins based on their overall spectral slope and the depth of the putative, partial ~0.9-μm absorption feature visible at the long wavelength end of the SDSS five-color spectrum.

Nine classes were defined by Carvano et al. (2010), including $D_p$, $L_p$, $X_p$, $C_p$, $A_p$, $S_p$, $Q_p$, $O_p$, and $V_p$. The subscript '*p*' is included to recognize the photometric basis of this classification system (Carvano et al., 2010). These classifications were applied to asteroids in Hasselmann et al. (2012). The $V_p$ class has the deepest putative absorption feature in the ~0.9-μm region and, as in other taxonomic systems, is generally thought to be consistent with basaltic-rich asteroids that are either directly associated with ejecta from (4) Vesta or represent basaltic material from other parent bodies throughout the main belt (Tholen, 1984; Bus and Binzel, 2002; Carvano et al., 2010). A sub-sample of the asteroids observed multiple times by SDSS were placed into different classes by the Carvano et al. (2010) system. As it concerns the $V_p$ class, multiple observations of the same asteroid tend to also be classified in the $Q_p$ and $S_p$ groups, due to their overall spectral similarity to the $V_p$ class and the presence of an ~0.9-μm absorption (Carvano et al., 2010). This ambiguity, which is expressed by Carvano et al. (2010) as a multiple taxonomic classification (i.e., $S_pV_p$, $S_pQ_pV_p$, etc.), highlights the strengths and weaknesses of taxonomic systems. Taxonomies are useful at providing broad classifications for a large number of objects, but sometimes fail the presumed taxonomic-compositional association upon detailed VNIR spectroscopic scrutiny of an individual asteroid (Gaffey et al., 1993; Hardersen et al., 2005, 2011).

Mainzer et al. (2012) combined WISE-derived albedos from Masiero et al. (2011) with the Carvano et al. (2010) $V_p$ taxonomic class to provide an additional constraint on the nature of these asteroids. As the $V_p$ class asteroids exhibits large albedos, this provides stronger evidence that this taxonomic class is associated with basaltic asteroids (Mainzer et al., 2012). The $V_p$ population of 650 asteroids has the largest median geometric albedo, $p_v$, of 0.343 ± 0.004 with a sample standard deviation of 0.105 and an albedo range from 0.047 to 0.771 (Mainzer et al., 2012). Thus, the $V_p$ class has both the bluest SDSS (*i'-z'*) magnitudes, which corresponds with the deepest putative ~0.9-µm absorption feature, and the largest median albedos of the Carvano et al. (2010) taxonomic classes.

### *1.3.3. $V_p$-type asteroid orbital distribution*

As a class, the $V_p$-type asteroids are predominantly clustered in orbital space near (4) Vesta ($a$ = 2.362 AU; $e$ = 0.090; $i$ = 7.134°: http://ssd.jpl.nasa.gov/?horizons) with a much smaller population of $V_p$-type asteroids located at larger semimajor axes and at higher inclinations (Carvano et al., 2010). Figure 1 displays the distribution of the 650 $V_p$-type asteroids in Mainzer et al. (2012) in a plot of inclination vs. semimajor axis, similar to that shown in Carvano et al. (2010). As there are 16,316 Vesta dynamical family members (Nesvorny, 2012) and most of the identified $V_p$-type asteroids are located in the orbital space near Vesta (see Figure 1), there is a significant likelihood of an association between Vesta dynamical family members and the $V_p$ asteroids from Mainzer et al. (2012). Mainzer et al. (2012) reinforce this suggestion by displaying the distribution of 650 $V_p$-type asteroids as a function of semimajor axis, eccentricity, and inclination (Figure 5: Mainzer et al., 2012). This figure includes bounding regions for the Nesvorny (2012) Vesta family, which shows that a large majority of the $V_p$-type asteroids exist in orbital space near (4) Vesta.

Evidence slowly continues to accumulate, however, that not all V-type asteroids are associated with (4) Vesta (Lazzaro et al., 2000; Hardersen et al., 2004; Roig et al., 2008; Duffard and Roig, 2009). The primary arguments against an origin from (4) Vesta are dynamical and mineralogical, and it is likely that these will be the two primary ways to try to identify basaltic remnants in the main asteroid belt that originate from other, now disrupted, parent bodies. Those $V_p$-type asteroids in Carvano et al. (2010) and Mainzer et al. (2012) that occupy orbital space beyond the Nesvorny (2012) Vesta dynamical family are most likely to originate from a parent body other than (4) Vesta, most notably if the candidate basaltic asteroids are beyond the 3:1 mean-motion resonance and at significantly different semimajor axes, eccentricities, and inclinations.

## 2. Observations and data reduction

### *2.1. Observations*

Observations of the eight WISE-defined $V_p$-type asteroids were made at the NASA Infrared Telescope Facility (IRTF), Mauna Kea, Hawai'i, from ~0420 to ~1515 UT on January 14, 2013 UT. The SpeX medium-resolution spectrograph was operated in prism mode to obtain asteroid, extinction star, and solar analog star spectra from ~0.7 to ~2.5-μm (Rayner et al., 2003, 2004). Extinction star observations are required for telluric corrections and solar analog star observations are necessary to correct for the overall NIR spectral slope when a non-G2V star is used as an extinction star (Hardersen et al., 2005). Weather conditions were photometric throughout the night with an average atmospheric seeing of ~0.83 and with local relative humidity ranging from ~14-36%.

Spectra were obtained utilizing the ...*ABBA*... pattern of observations that alternates the position of the target within the spectrograph slit during a set of observations. This technique allows for easier background sky subtraction during the data reduction process. Observations also utilized the 0.8" x 15" slit and all spectra were obtained at or near the parallactic angle to maximize light throughput to the detector.

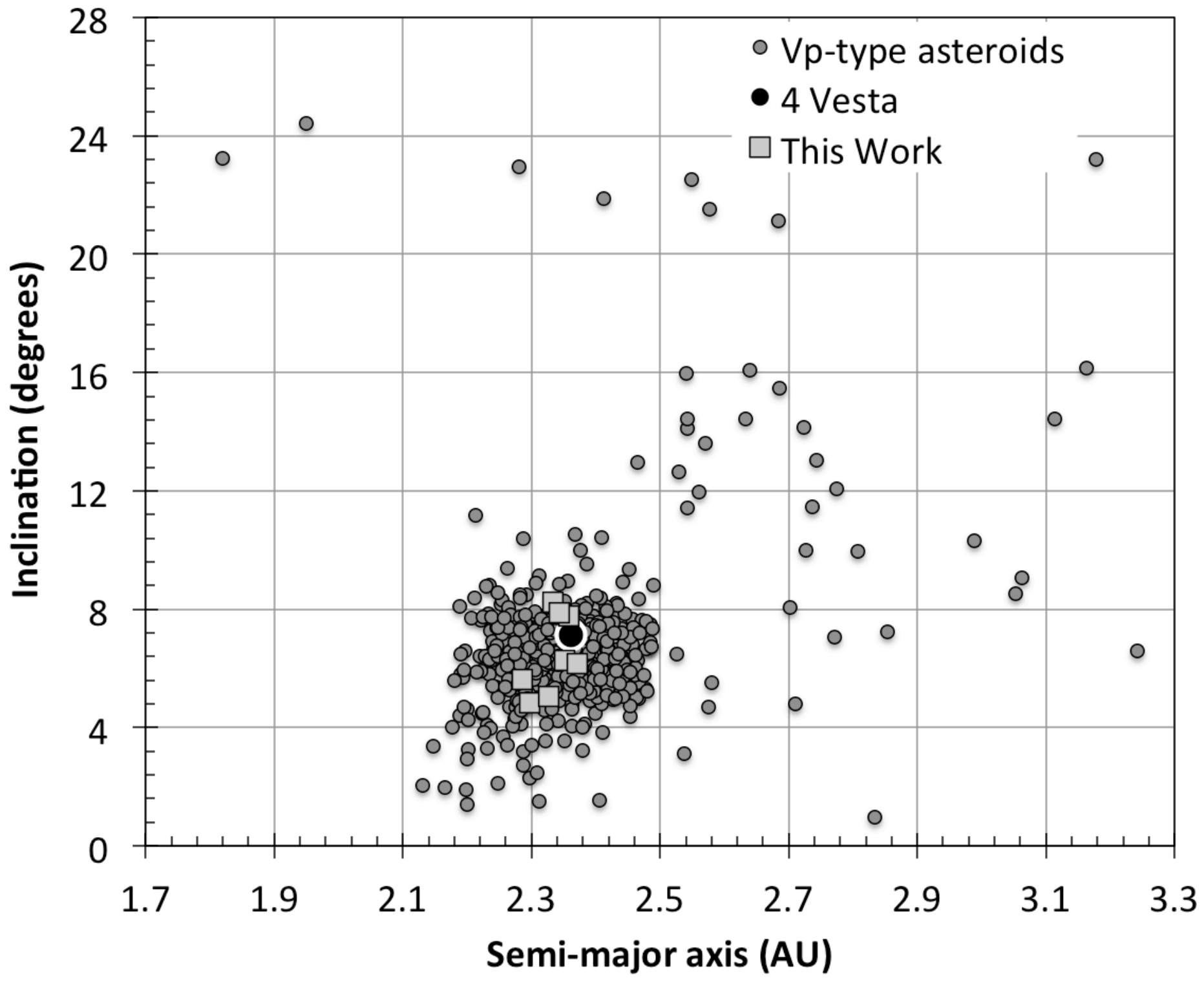

Figure 1. The inclination vs. semi-major axis distribution for 650 $V_p$-type asteroids, along with (4) Vesta. Most of the $V_p$-type asteroids cluster in orbital space near (4) Vesta and include the eight asteroids in this work. $V_p$-type asteroid data provided by A. Mainzer. Vesta dynamical data obtained from the JPL Horizons ephemeris service at: http://ssd.jpl.nasa.gov/?horizons.

Table 1 displays the specific objects observed during this night. All of the asteroids were relatively faint with apparent visual magnitudes ranging from ~16.4 to ~17.3. Total integration times for each asteroid ranged from 1200-4080 seconds, corresponding to a total number of spectra per asteroid ranging from 10-34. All individual asteroid spectrum integration times were 120 seconds. Each asteroid spectrum in a set of observations is examined during the data reduction process to determine overall spectral quality and the effectiveness of the telluric feature corrections. Poor quality spectra are omitted in the averaging process to increase the quality of the average asteroid spectrum.

| Asteroid | Observation date (UT) | Apparent *V* magnitude | Phase angle | Total int. time (sec) | Total spectra | Extinction star | Solar analog star |
|---|---|---|---|---|---|---|---|
| (3867) Shiretoko | 1/14/13 | 16.72 | 24.3° | 4080 | 34 | HD 106210 (G3V) | HD 28099 (G2V) |
| (5235) Jean-Loup | 1/14/13 | 16.54 | 10.1° | 1200 | 10 | HD 82159 (G9V) | HD 28099 (G2V) |
| (5560) Amytis | 1/14/13 | 17.28 | 10.6° | 1200 | 10 | TYC 1867-2584-1 (G5V) | HD 28099 (G2V) |
| (6331) 1992 FZ1 | 1/14/13 | 16.37 | 1.2° | 1200 | 10 | HD 59374 (F8V) | HD 28099 (G2V) |
| (6976) Kanatsu | 1/14/13 | 16.51 | 10.2° | 2160 | 18 | TYC 1867-217-1 (G0V) | HD 28099 (G2V) |
| (17469) 1991 BT | 1/14/13 | 16.67 | 16.9° | 1200 | 10 | HD 88725 (G1V) | HD 28099 (G2V) |
| (29796) 1999 CW77 | 1/14/13 | 16.76 | 10.9° | 1200 | 10 | HD 254085 (G0V) | HD 28099 (G2V) |
| (30872) 1992 EM17 | 1/14/13 | 16.66 | 0.9° | 1200 | 10 | HD 59374 (F8V) | HD 28099 (G2V) |

Table 1. Details for eight WISE-defined $V_p$-type asteroids observed on January 14, 2013 UT, at the NASA Infrared Telescope Facility (IRTF), Mauna Kea, Hawai'i, using the SpeX near-infrared spectrograph (Rayner et al., 2003, 2004). Observations were made in the SpeX low-resolution (i.e., prism) mode at the parallactic angle, and using the 0.8" slit with an effective resolution of $R$ ~ 95. For this observing run, Mauna Kea skies were clear with the local relative humidity ranging from ~14-36%. Atmospheric seeing averaged ~0.83".

Each asteroid was paired with an F-/G-type main-sequence star that is located near the position of the asteroid on the sky, i.e., within a few degrees. This close proximity of the asteroid-star pair increases the likelihood that atmospheric conditions were similar for both objects during each set of observations. This observing strategy is effective for the later removal of spectral telluric absorptions during the data reduction process. Thirty spectra of solar analog star HD 28099 were obtained near meridian passage, which are used to correct for spectral slope variations during the data reduction process when extinction star spectral types are non-solar, i.e., non-G2V.

### *2.2. Data reduction*

All data were reduced using Spextool, a collection of IDL routines that perform spectral calibrations, telluric corrections, channel shifts, averaging, and display functions (Cushing et al., 2004). Argon arc spectra, acquired during the observing run, are used by Spextool to perform a wavelength calibration that converts CCD channels to wavelength values. Multiple series of flat field images obtained at the telescope were also median-combined and applied to all object spectra to correct for CCD pixel sensitivity variations and any non-uniformity in array illumination.

Effective spectral telluric corrections require two sets of extinction star observations per asteroid across an airmass range that bound the airmass values at which the asteroid spectra were obtained. Both sets of stellar spectra (i.e., typically 10 spectra per set) are then averaged. A ratio of each asteroid spectrum to the extinction star average is then performed, which includes a channel offset correction to compensate for the raw object flux landing on slightly different sets of pixels during successive integrations. This channel offset effect is caused by telescope flexure and loading when pointing at different positions in the sky throughout the night. Asteroid spectra with poor quality telluric corrections or other obvious artifacts are omitted before producing an average asteroid spectrum.

The atmosphere- and channel-corrected asteroid spectra are then averaged to produce a preliminary normalized average asteroid spectrum. Each solar analog star spectrum (i.e., in this case, HD 28099) is reduced in the same manner as each asteroid spectrum. An average spectrum for HD 28099 is then created for each extinction star to account for each extinction star's spectral slope variance from the nominal G2V spectral class and association with a particular asteroid. The only difference is that telluric corrections for the solar analog star are smoothed to remove residual telluric absorptions. This step is performed because solar analog stars are usually in different locations in the sky compared to the extinction stars, which makes correcting for telluric absorptions difficult. Spectral smoothing is also justified because broad mineral or water vapor absorptions are not present in the NIR region in stellar spectra (Hardersen et al., 2005). The final step is to ratio each average asteroid spectrum to its average solar analog spectrum, which occurs outside of Spextool in programs such as Microsoft Excel or in IDL. The overall reduction process is summarized below:

*Avg. Normalized Asteroid Spectrum = (Avg. Asteroid Spectrum/Avg. Extinction Star Spectrum) / (Avg. Solar Analog Spectrum / Avg. Extinction Star Spectrum)*

*2.3. Data analysis*

Data analysis of the resulting asteroid spectra involves two steps: 1) characterizing the relevant band parameters, and 2) applying laboratory-based mineral and meteorite calibrations that allow interpretations to constrain an asteroid's geologic nature and suggest any possible meteorite affinities. Band parameters characterized in this work include the ~0.9-μm (i.e., Band I) center, the ~1.9-μm (i.e., Band II) center, absorption band area and the associated BAR (BAR: Band II Area/Band I Area), and the absorption band depth. The first step involves isolating each absorption feature from the average spectrum by removing the overall continuum, which eliminates any inherent slope that is present in the average spectrum. Each absorption feature is extracted from the average asteroid spectrum by extending a linear continuum across the absorption feature and taking the ratio of the reflectance of the linear continuum to the absorption feature. The linear continuum extends across the absorption feature and terminates at the local maximum on each side of the feature. While a linear continuum does not accurately reflect the actual continuum across the spectrum, the same continuum-removal process is applied to mineral and meteorite NIR spectra to allow direct comparisons of the resulting band centers. Band centers and band areas are then calculated for each continuum-removed absorption feature.

Band centers are defined as the minimum reflectance of a continuum-removed absorption feature. Measuring band centers in this manner, as compared to measuring the band minimum for a non-continuum-removed absorption feature, is directly related to crystal field theory and the absorptions produced by the relevant transition metal ions (i.e., $Fe^{2+}$) within mafic silicate minerals (Adams, 1974; Gaffey, 1976; Burns et al., 1993a). Absorption band centers for orthopyroxenes and Type B clinopyroxenes display a systematic change in band center with increasing abundances of $Ca^{2+}$ and $Fe^{2+}$ within the mineral crystal structures (Adams, 1974; Gaffey, 1976). This trend is effective in differentiating among the HED meteorite analogs as the band centers for the diogenites, howardites, and eucrites shift to longer wavelengths due to the progressively larger abundances of $Fe^{2+}$ in the mineral crystal structures (Gaffey, 1997).

Band Area is defined as the area of a continuum-removed absorption feature. Band Area Ratio (BAR) is defined as the ratio of the areas of the Band II absorption feature to the Band I absorption feature (BAR = Band II Area/Band I Area: Cloutis et al., 1986; Gaffey et al., 2002). Accurate calculation of band areas is critically dependent upon measuring the entire absorption feature. For the Band I feature, this requires that the short-wavelength rollover is present that defines the short wavelength end of the feature. The long-wavelength end of the Band I feature is usually defined as the local maximum that corresponds to the short-wavelength end of the Band II absorption feature (Cloutis et al., 1986; Gaffey et al., 2002). This is the case for meteorite and asteroid NIR spectra where olivine and pyroxene are the major mafic silicates present. For Band II, the critical issue is capturing the entire feature at the long-wavelength end of the spectrum. Due to the breadth of the Band II pyroxene feature in V-type asteroid spectra, it is usually not possible to capture this entire feature because it extends to wavelengths beyond that present in the available data. A consequence of this is that the reported BARs for most V-type asteroids will be somewhat smaller compared to the BAR if the entire feature had been measured.

One characteristic of V-type asteroids and orthopyroxene/Type B clinopyroxene NIR spectra that mitigates this problem is their relatively large BAR values that exceed BAR values for NIR spectra of other mineral mixtures and meteorites. BAR values for V-type asteroids are also consistent with BAR values from HED meteorite spectra (Gaffey et al., 1993; Hardersen et al., 2004). Another factor that influences band area calculations and their BAR values is the inherent quality of the NIR spectrum and the telluric corrections at ~1.4- and ~1.9-μm. Noisier spectra will produce greater variations in band area measurements, which will increase the associated errors in the resulting BAR calculations.

Absorption band depth is another spectral band parameter that is measured, which is defined as the fractional depth of the continuum-removed absorption feature. It is reported as a percentage of depth relative to the normalized continuum. Observational and physical factors that affect absorption band depth include the phase angle of the observations, the type and relative abundance(s) of the surface minerals present, the presence of absorbing transition metal ions in the mineral crystal structures, the presence and relative abundance of opaque minerals and organic compounds in the surface materials, and the particle sizes and particle size distribution (Gradie and Veverka, 1986; Cloutis et al., 1990; Burns, 1993a; Benner et al., 2008; Sanchez et al., 2012).

*2.3.1. MATLAB data reduction*

A series of MATLAB routines from Reddy et al. (2011) were employed to derive spectral band parameters for the 8 $V_p$-type asteroids in this work. These routines calculate band centers, band areas, BARs, and band depths, as defined and described above. Text files for each average asteroid spectrum, including the wavelength and normalized reflectance values, were created and imported into MATLAB for the analysis.

Two iterations of band center calculations were performed: 1) using a 3$^{rd}$-order polynomial fit, and 2) using a 5$^{th}$-order polynomial fit. For each continuum-removed absorption feature, a polynomial was fit to a spectral region that bounds the wavelength position of the minimum absorption. The portions of each feature chosen at the short- and long-wavelength sides were varied to determine how these choices affected the band center calculations. At least 10 fits were conducted for each polynomial order to constrain the uncertainties for each absorption band center. The band center uncertainties represent the maximum deviations from the mean of all of the band center fits.

*2.3.2. SARA data reduction*

A second series of spectral band measurements was calculated using the 'Spectral Analysis Routine for Asteroids', or SARA, which is a collection of IDL routines that automatically calculate absorption band centers, band areas, BARs, band slope, and band depths from an input text file of an average normalized asteroid reflectance spectrum (Lindsay et al., 2013, 2014). As compared to the MATLAB routines where the spectral band parameters are obtained semi-automatically one set at a time, SARA inputs a spectrum and calculates all of the spectral band parameters simultaneously.

SARA isolates absorption features using a linear continuum in a manner consistent with the MATLAB routines and as described above. SARA calculates absorption band centers using 3$^{rd}$, 4$^{th}$, and 5$^{th}$ order polynomials, which produce three sets of results per run for each spectral band parameter (i.e., band center, band area, BAR, band depth, band slope). Multiple runs of SARA using the same asteroid spectrum produced very similar or identical results, verifying the consistency of the program in extracting and calculating the various band parameters. SARA calculates band areas, and the resulting BAR, via the trapezoidal rule.

Comparison of the derived MATLAB and SARA spectral band parameters show very consistent results for band centers, band areas, BARs, and band depths. The results reported in Table 2 represent averages of the MATLAB and SARA results and include the maximum band center and BAR uncertainties when considering all of the analyses together.

*Section 2.3.3. Temperature Corrections*

The derived band centers for the eight $V_p$-type asteroids were temperature-corrected using the protocols from Burbine et al. (2009) and Reddy et al. (2012c). Existing WISE albedos for each asteroid, as well as each asteroid's heliocentric distance at the time of observation, and other input variables (solar luminosity, beaming factor =

1.00, infrared emissivity = 0.90), were used to estimate the surface temperature for each asteroid (Burbine et al., 2009).

Estimated surface temperatures and raw band centers were input into the HED temperature correction algorithms of Reddy et al. (2012c). The results produce a band shift, in μm, that is added to each raw band center to compensate for the temperature differential between the asteroid's surface and the room temperature environment (~300 K) where laboratory spectra of meteorites are measured. The temperature-corrected band centers were subsequently used to estimate the average surface pyroxene chemistry for the asteroids.

## 3. Results and interpretations

### *3.1. Average asteroid NIR reflectance spectra*

The final, average NIR reflectance spectrum for each of the eight $V_p$-type asteroids is shown in Figures 2(A-H). All of the average spectra exhibit variably deep absorption features in the ~0.9- and ~1.9-μm spectral regions that are typical of asteroid surfaces dominated by pyroxene (Adams, 1974; Singer, 1981; Cloutis and Gaffey, 1991; Schade et al., 2004). The ~0.9-μm absorption feature for each average spectrum exhibits uniformly high signal-to-noise ratio (SNR); all the spectra extend to the short-wavelength maxima of the absorption features at ~0.70-0.75-μm. The ~0.9-μm feature for (5560) Amytis is the only feature in this spectral region among all eight asteroids that does not exhibit a symmetrical shape due to the kink near the minimum of the feature.

Spectral telluric corrections are generally effective with only minor residual noise present in the ~1.32-1.40-μm region for most of the asteroids. Additional point-to-point data scatter and/or distortion of the spectra are variably present near the minima of the ~1.9-μm absorption features from ~1.80-2.05-μm, most notably for the average spectrum of (6331) 1992 FZ1 and (30872) 1992 EM17. Data scatter is also apparent for some asteroids at the long-wavelength end of the spectrum beyond ~2.4-μm, including (5560) Amytis, (6331) 1992 FZ1, (29796) 1999 CW77, and (30872) 1992 EM17.

With the exception of (6331) 1992 FZ1, all of the average spectra exhibit an inflection in the ~1.2-μm spectral region. Inflections and weak absorptions in this region have previously been attributed to plagioclase feldspar, Fe-rich orthopyroxene, and olivine (Hardersen et al., 2006, and references therein); this inflection is also present in laboratory spectra of eucrites and howardites (Gaffey, 1976). Olivine can be ruled out as a possible cause for this inflection due to its absence as a major mineral within HED meteorites and their NIR spectra.

### *3.2. Spectral band analysis: a three-tiered test*

A three-tiered spectral band analysis test is applied to each $V_p$-type asteroid spectrum to determine the likelihood that each asteroid is: 1) similar to the HED meteorites; and 2) associated with a particular member of the HED meteorites (i.e., howardites, eucrites, diogenites). Absorption band centers, band areas, and band depths are calculated as described in Section *2.3*. Table 2 reports the derived band centers, Band Area Ratios (BAR), and band depths for each asteroid. Table 3 reports the derived pyroxene chemistries from two different calibrations (Gaffey et al., 2002; Burbine et al., 2009), plus their average, that will be discussed in Section *3.2.2.*

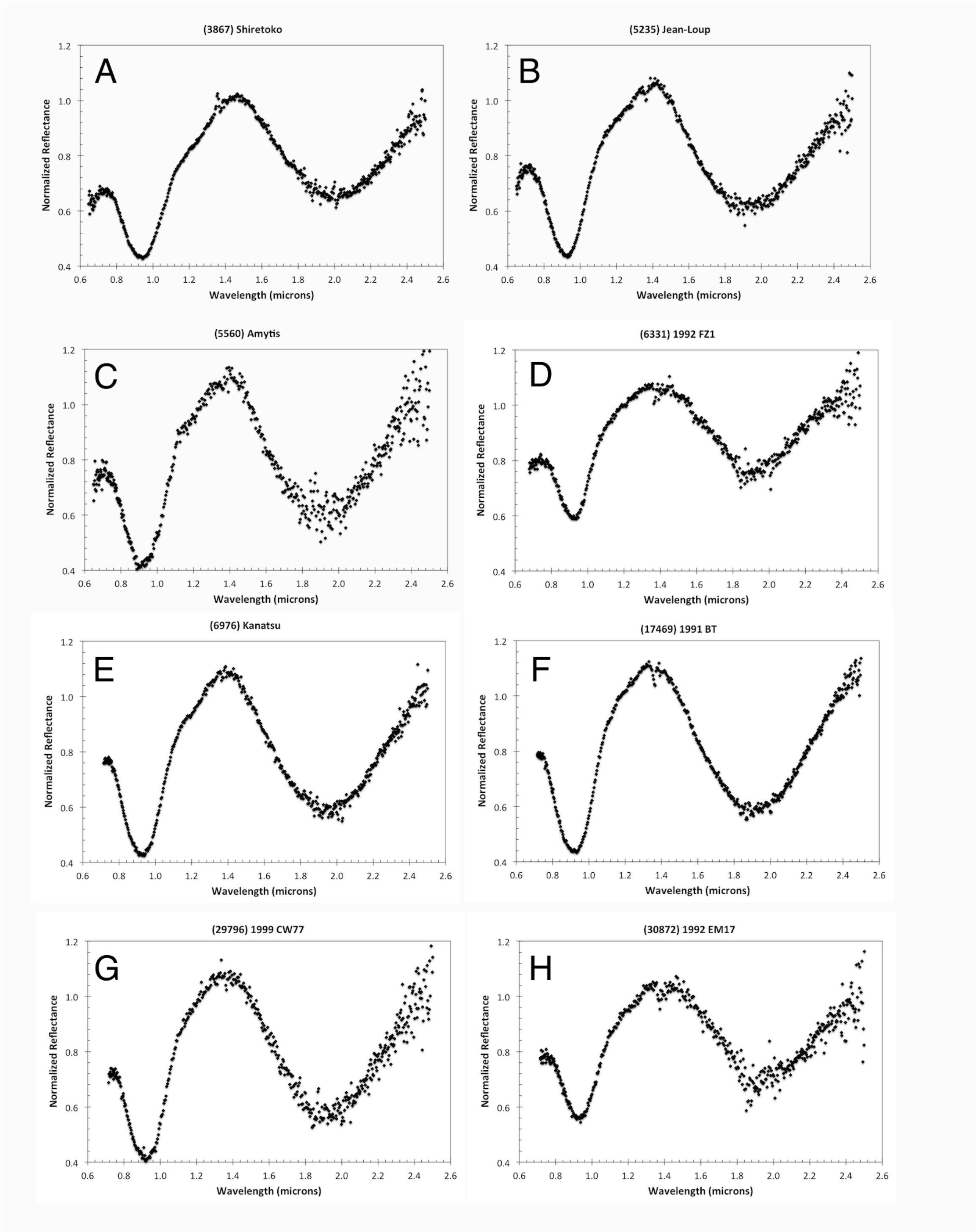

Figure 2(A-H). Average near-infrared (NIR) reflectance spectra for eight WISE-defined $V_p$-type asteroids: A) a 34-spectrum average for (3867) Shiretoko; B) a 10-spectrum average for (5235) Jean-Loup; C) a 10-spectrum average for (5560) Amytis; D) a 10-spectrum average for (6331) 1992 FZ1; E) an 18-spectrum average for (6976) Kanatsu; F) a 14-spectrum average for (17469) 1991 BT; G) a 6-spectrum average for (29796) 1999 CW77; and H) a 10-spectrum average for (30872) 1992 EM17. The spectra have been normalized to unity at 1.5 μm.

*3.2.1. Band I vs. BAR analysis*

Cloutis et al. (1986) reported results from a laboratory calibration that showed variations of spectral Band I centers and BARs with varying proportions of olivine and orthopyroxene in a two-mineral-only mixture. Gaffey et al. (1993) adopted this technique and defined general regions in the Band I vs. BAR plot for olivine-rich objects, ordinary chondrites, and basaltic achondrites. This was the basis for the Gaffey et al. (1993) S-type asteroid sub-classification system, which included seven sub-types (i.e., S-I to S-VII). The basaltic achondrite (BA) region is characterized by large BAR values ($1.5 \leq$ BAR $\leq$ 2.7), which corresponds to a large proportion of orthopyroxene compared to olivine. The BA zone is adjacent to, and just above, the S-VII sub-types, where the S-VII zone includes pyroxene-rich asteroids that may have meteoritic affinities different from the HED meteorites (Gaffey et al., 1993). For example, previous work by Gaffey (1997) and Hardersen et al. (2004) place (4) Vesta and (1459) Magnya, respectively, in (or beyond) the BA region that is interpreted as asteroids with HED- or HED-like surface mineralogies.

Figure 3 shows a Band I center vs. BAR plot that includes data for the eight WISE-defined $V_p$-type asteroids in this paper. As shown in Figure 3, seven of the eight asteroids plot securely within the BA zone, while (3867) Shiretoko bounds the lower BAR boundary of the BA zone. This result indicates that these eight asteroids are broadly consistent with a basaltic achondrite interpretation based on the spectral parameters in the Band I vs. BAR plot.

| Asteroid | Band I (μm) | Band II (μm) | BAR | Band I depth (%) | Band II depth (%) |
|---|---|---|---|---|---|
| (3867) Shiretoko | 0.951 ± 0.005 | 2.008 ± 0.005 | 1.54 ± 0.04 | 41 ± 9 | 32 ± 1 |
| (5235) Jean-Loup | 0.936 ± 0.004 | 1.930 ± 0.005 | 1.94 ± 0.08 | 46 ± 8 | 38 ± 3 |
| (5560) Amytis | 0.932 ± 0.011 | 1.934 ± 0.010 | 1.98 ± 0.12 | 50 ± 5 | 42 ± 1 |
| (6331) 1992 FZ1 | 0.938 ± 0.003 | 1.928 ± 0.012 | 2.06 ± 0.09 | 31 ± 9 | 27 ± 5 |
| (6976) Kanatsu | 0.939 ± 0.003 | 1.967 ± 0.004 | 2.09 ± 0.11 | 49 ± 4 | 44 ± 2 |
| (17469) 1991 BT | 0.933 ± 0.005 | 1.926 ± 0.005 | 2.40 ± 0.01 | 48 ± 6 | 46 ± 3 |
| (29796) 1999 CW77 | 0.937 ± 0.006 | 1.930 ± 0.006 | 2.16 ± 0.15 | 49 ± 7 | 45 ± 2 |
| (30872) 1992 EM17 | 0.936 ± 0.008 | 1.933 ± 0.017 | 2.25 ± 0.20 | 33 ± 7 | 31 ± 6 |

Table 2. Derived spectral band parameters for eight WISE-defined $V_p$-type asteroids. Band centers, band area ratios, and band depths are reported for continuum-removed absorption features. Band centers are averages of five sets of polynomial fitting attempts using two analysis techniques. BARs are the averages from three separate sets of calculations with the exception of (17469) 1991 BT, which is an average of two sets of calculations. Band center and BAR areas conservatively represents the maximum deviation of an individual measurement from the average. Band depth represents the fraction that an absorption feature extends from a normalized linear continuum (i.e., 1) to 0, as in Band Depth = 1 – BCD, where band center depth (BCD) is the minimum normalized position of the absorption band center.

### *3.2.2. Pyroxene band-band plot analysis*

Adams (1974) originally demonstrated that pyroxene absorption band centers vary systematically as a function of their $Ca^{2+}$ and $Fe^{2+}$ content. Band I and Band II centers are plotted for 55 HED meteorites from Le Corre et al. (2011), along with the same parameters for the asteroids in this study. Figure 4 displays a pyroxene band-band plot that includes spectral data for the eight WISE-defined $V_p$-type asteroids and 55 HED meteorites (Le Corre et al., 2011).

Figure 4 shows that (3867) Shiretoko displays band centers consistent with a eucrite interpretation as it is located near and just above the eucrite region on the right side of the figure. The plot of the data for (6976) Kanatsu suggests a howardite or eucrite-rich howardite composition based on its location near the intersection of the eucrite and howardite regions.

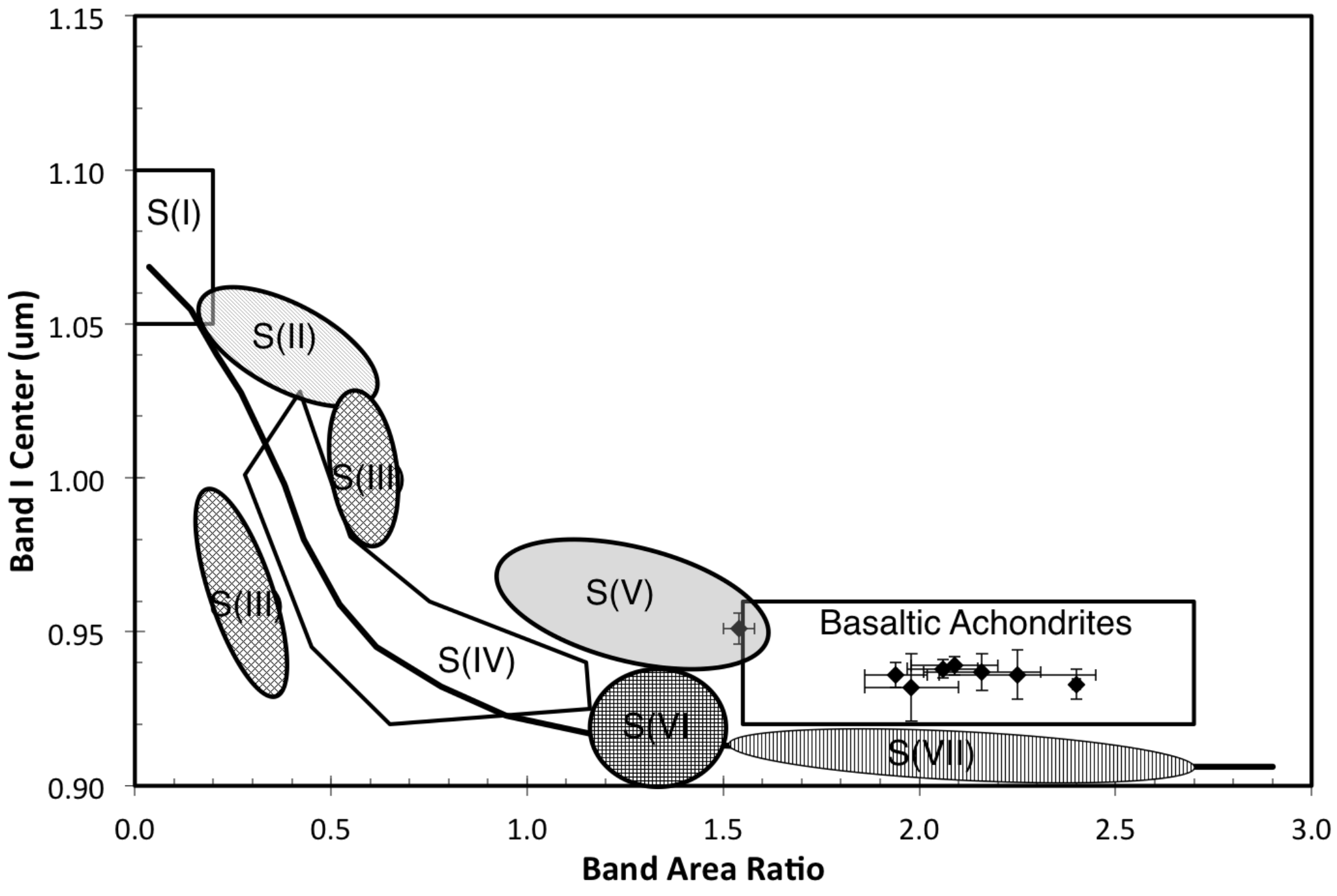

Figure 3. Band I center vs. BAR plot for eight WISE-defined $V_p$-type asteroids. The data for the eight asteroids, plus error bars, are found on, or within, the basaltic achondrite (BA) zone, as defined by Gaffey et al. (1993). The solid line from upper left to lower right is an olivine-orthopyroxene mixing line where orthopyroxene abundance increases as BAR increases.

The remaining six asteroids are clustered together in the howardite region to the upper right of the diogenite zone. This suggests a dominantly howardite-like composition, possibly with an enhancement in the diogenite component. All eight asteroid band-band data plot slightly above the HED data, which suggests a possible minor contribution from olivine or a high-Ca pyroxene phase. These minerals (olivine, Type A clinopyroxene) only exhibit Band I absorption features; however, Band I centers for olivine will increase with increasing relative abundances while no band center trend is seen for Type A clinopyroxenes (King and Ridley, 1987; Burns, 1993a; Schade et al., 2004; Sanchez et al., 2014). However, any contribution from these phases will be minor as the Figure 4 data plots only slightly above the HED data and a significant contribution from another mineral phase is not apparent in Figure 3 or Figure 5. The asteroid data in the band-band plot are consistent with an affinity to the HED meteorites and, as noted, with particular associations among the eucrite and howardite members of the HED group.

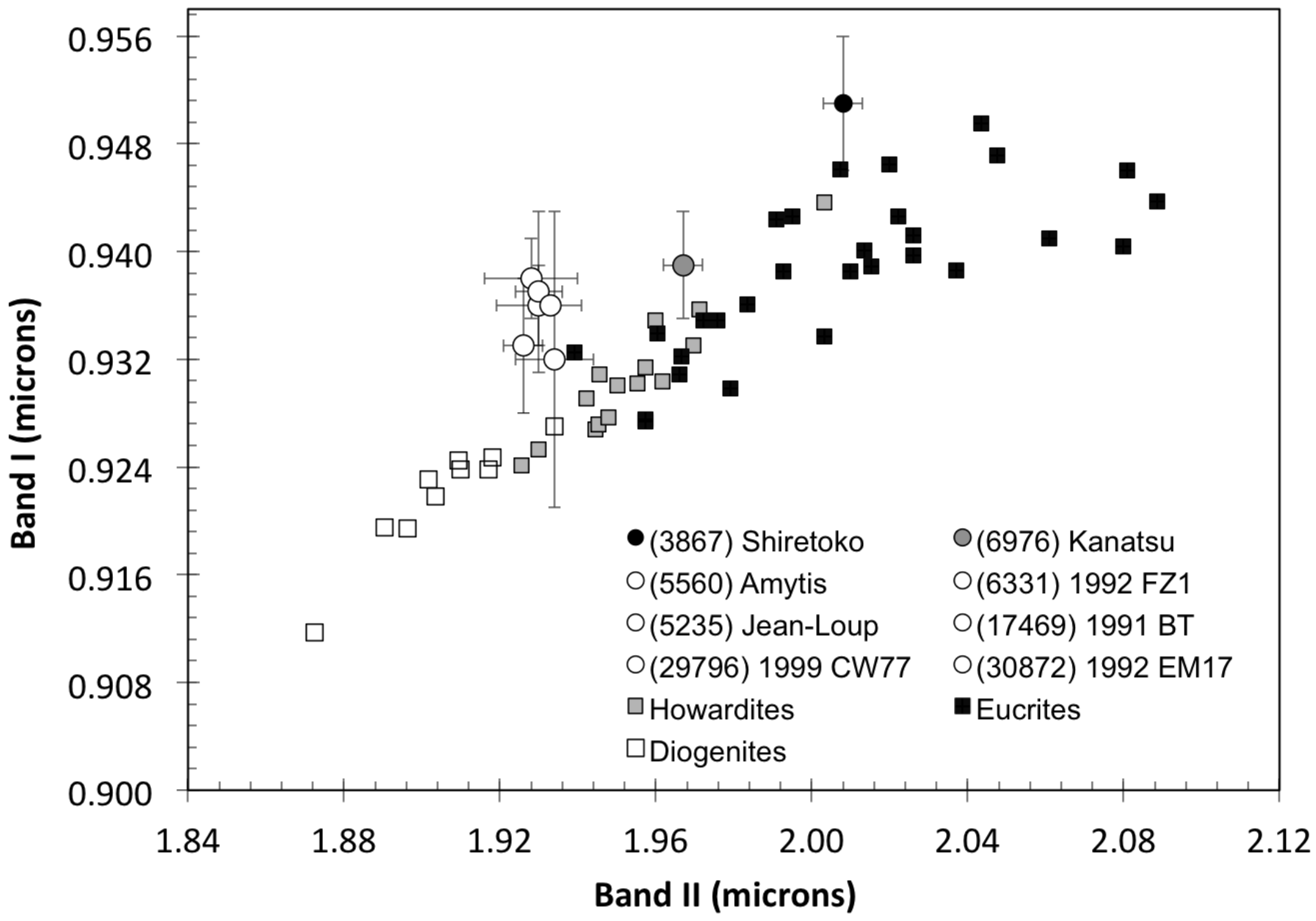

Figure 4. Pyroxene band-band plot for eight WISE-defined $V_p$-type asteroids (circles) and a sampling of HED meteorites (squares) from Le Corre et al. (2011). Diogenites are open squares, howardites are gray squares, and eucrites are black squares. Open circles represent six $V_p$-type asteroids that display a howardite composition with a possible enhancement in the diogenite component. The gray circle represents (6976) Kanatsu, which exhibits a howardite composition with a possible enhancement in the eucrite component. The black circle represents (3867) Shiretoko, which exhibits a eucritic composition.

### *3.2.3. Pyroxene quadrilateral analysis*

The band-band analysis in Section *3.2.2* can be extended to estimate the average surface pyroxene chemistries for the eight WISE-defined $V_p$-type asteroids, which can be compared to typical pyroxene chemistries found for the HED meteorites. Asteroid Band I and Band II centers, which were temperature-corrected, were used to estimate average surface pyroxene chemistries for each asteroid using the techniques in Gaffey et al. (2002) and Burbine et al. (2009). For Gaffey et al. (2002), the single-pyroxene option was used, which is appropriate for basaltic/HED-like asteroids whose major mineralogy includes a single pyroxene. The Burbine et al. (2009) technique is directly applicable to basaltic/HED-like asteroids. Results from these two techniques, plus their averages, are shown in Table 3.

The average surface pyroxene chemistries for each asteroid in Table 3 were then plotted within a pyroxene quadrilateral to compare with pyroxene chemistries derived from a sample of HED meteorites. HED pyroxene chemistries were calculated from HED spectral band parameters by Le Corre et al. (2011), which were derived from the original HED/pyroxene data in Mittlefehldt et al. (1998). Figure 5 shows data for the eight asteroids and HED meteorite zones. (3867) Shiretoko plots to the right of the other asteroids due to its higher Fs pyroxene abundance and its spatial association with the eucrites. The other seven asteroids are clustered together in a relatively small region of the quadrilateral, which reflects their similar pyroxene chemistries. Figure 42 in Mittlefehldt et al. (1998) displays a pyroxene quadrilateral with ferrosilite (Fs) chemistry data for basaltic eucrites, cumulate eucrites, diogenites, and the meteorite, Pomozdino. The basaltic eucrites have a range of Fs values from $Fs_{\sim 40-60}$. Data for the diogenites, which are more magnesian in their pyroxene content, display a range of Fs values from $Fs_{\sim 20-33}$ and are less calcic than the eucrites. The cumulate eucrites plot at intermediate Fs values that span the range between the diogenite and basaltic eucrite data. The data for (3867) Shiretoko are consistent with the basaltic eucrite data in Figure 42 of Mittlefehldt et al. (1998).

| Asteroid | Burbine et al. (2009) pyroxene chemistry | Gaffey et al. (2002) pyroxene chemistry | Average pyroxene chemistry |
|---|---|---|---|
| (3867) Shiretoko | $Wo_{14}Fs_{54}$ | $Wo_{16}Fs_{43}$ | $Wo_{15}Fs_{48}$ |
| (5235) Jean-Loup | $Wo_{8}Fs_{39}$ | $Wo_{11}Fs_{35}$ | $Wo_{10}Fs_{37}$ |
| (5560) Amytis | $Wo_{6}Fs_{36}$ | $Wo_{9}Fs_{36}$ | $Wo_{8}Fs_{36}$ |
| (6331) 1992 FZ1 | $Wo_{8}Fs_{39}$ | $Wo_{12}Fs_{38}$ | $Wo_{10}Fs_{39}$ |
| (6976) Kanatsu | $Wo_{10}Fs_{44}$ | $Wo_{13}Fs_{40}$ | $Wo_{12}Fs_{42}$ |
| (17469) 1991 BT | $Wo_{7}Fs_{37}$ | $Wo_{10}Fs_{34}$ | $Wo_{9}Fs_{35}$ |
| (29796) 1999 CW77 | $Wo_{8}Fs_{39}$ | $Wo_{11}Fs_{37}$ | $Wo_{10}Fs_{38}$ |
| (30872) 1992 EM17 | $Wo_{9}Fs_{42}$ | $Wo_{14}Fs_{39}$ | $Wo_{12}Fs_{40}$ |

Table 3. Derived average surface pyroxene chemistries for eight WISE-defined $V_p$-type asteroids. Chemistries were calculated using the techniques of Gaffey et al. (2002) and Burbine et al. (2009), along with the average of the two results. The results are broadly consistent within the error bars of the derived chemistries, which are $Wo_{(\pm 3\text{-}4)}$ and $Fs_{(\pm 4\text{-}5)}$ (Gaffey et al., 2002) and $Wo_{(\pm 1.1)}$ and $Fs_{(\pm 3.3)}$ (Burbine et al., 2009).

Pyroxene chemistry data for the remaining asteroids are located between the basaltic eucrite and diogenite data. The data for these eight asteroids are consistent with both the howardites and cumulate eucrites. The howardite interpretation is more likely due to the violent collisional and mixing environment on, and around, (4) Vesta that produced the Vestoids. Preserving large fragments of cumulate eucrites as ejecta is less likely than the dominant composition seen on Vesta, which is howarditic in nature. Also, the positions of six of the asteroids in Figure 4 are largely offset from the existing eucrite data, which makes a cumulate eucrite interpretation more difficult. Table 4 lists each $V_p$-type asteroid with its most likely HED meteorite analog.

### *3.2.4. Band depth analysis*

A portion of Table 2 displays the continuum-removed band depths for both the ~0.9- and ~1.9-μm absorption features for each asteroid. Most asteroids display band depths that typically range from ~40-50% of the normalized continuum. (6331) 1992 FZ1 and (30872) 1992 EM17, however, display band depths that are notably shallower than those for the other asteroids, by approximately one-third. Band depths for (3867) Shiretoko are somewhat shallower, especially the Band II feature, compared to most of the asteroids.

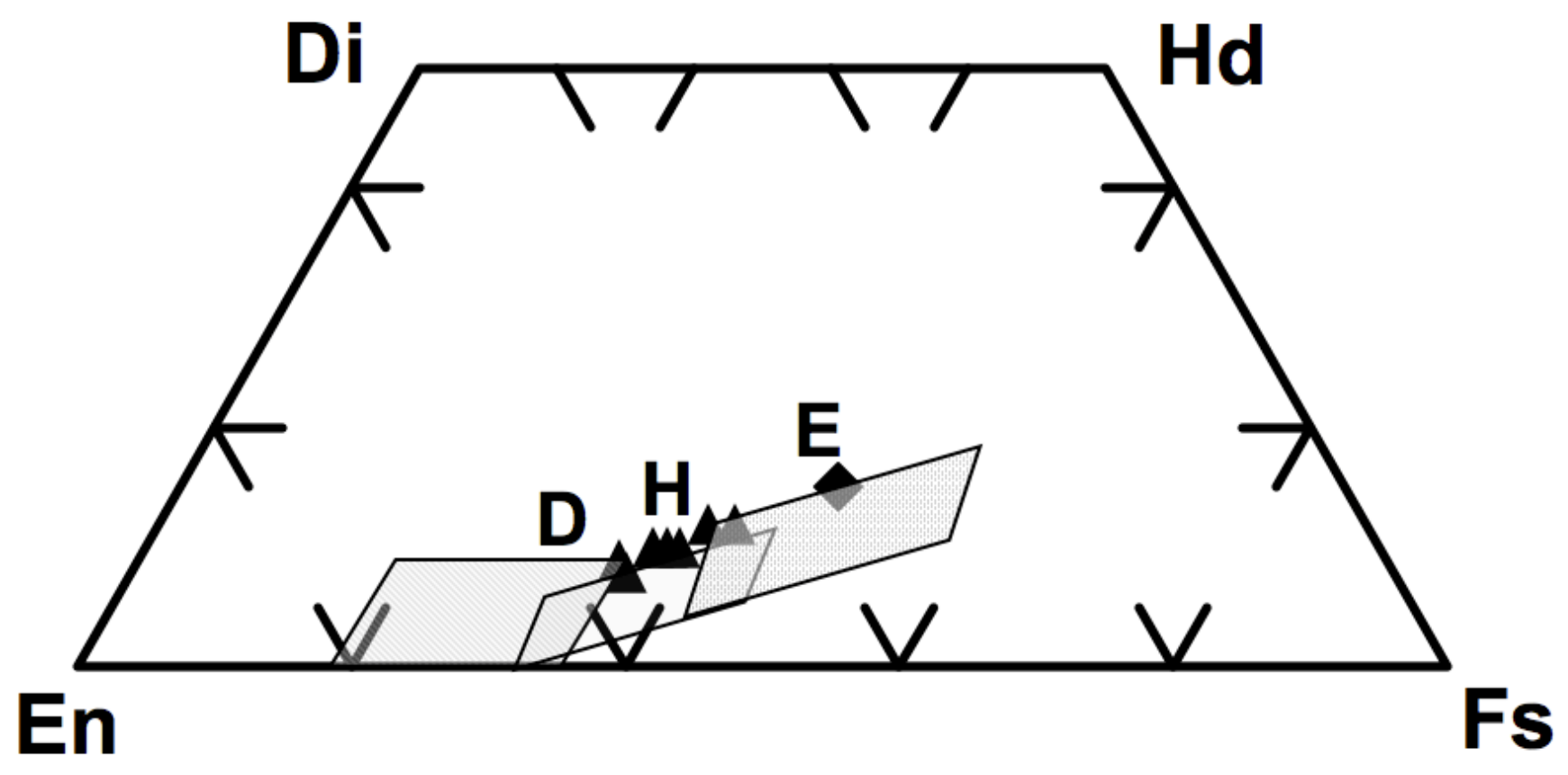

Figure 5. Pyroxene quadrilateral with data for the eight WISE-defined $V_p$-type asteroids and representative HED meteorite data from Le Corre et al. (2011). Asteroids represented by a solid diamond (3867 Shiretoko) and solid triangles. The quadrilaterals labeled as D, H, and E represent the bounds of the individual data points from Le Corre et al. (2011). D = diogenite zone, H = intermediate howardite zone, E = eucrite zone. En = enstatite, Fs = ferrosilite, Di = diopside, Hd = hedenbergite. Uncertainties for the plotted data are $Wo_{(\pm 3\text{-}4)}$ and $Fs_{(\pm 4\text{-}5)}$ (Gaffey et al., 2002).

| Asteroid | HED meteorite analog |
|---|---|
| (3867) Shiretoko | Eucrite |
| (5235) Jean-Loup | Howardite |
| (5560) Amytis | Howardite (diogenite-rich?) |
| (6331) 1992 FZ1 | Howardite |
| (6976) Kanatsu | Howardite (eucrite-rich?) |
| (17469) 1991 BT | Howardite (diogenite-rich?) |
| (29796) 1999 CW77 | Howardite |
| (30872) 1992 EM17 | Howardite |

Table 4. The eight WISE-defined $V_p$-type asteroids in this paper and their suggested HED meteorite analogs based on the existing evidence.

Sanchez et al. (2012) report that band depths for NIR spectra of Near-Earth Asteroids and ordinary chondrites increase with increasing phase angle. Asteroid Band I depths increase with increasing phase angles from ~2-70° and Band II depths increase with increasing phase angle from ~2-55° (Sanchez et al., 2012). Comparison of asteroid phase angles during the observations shown in Table 1 and their derived band depths in Table 2 show no correlation between the two variables. The eight WISE-defined $V_p$-type asteroids were observed across phase angles ranging from ~1-24°. Plots of Band I/II depth vs. phase angle for our small dataset show an increase in band depth for phase angles from ~1-11°, but then a decreasing band depth at larger phase angles.

Reddy et al. (2012b) reported the presence of dark material (DM) on Vesta in a variety of locations and geologic settings. Based on observational, spectral, and modeling analysis, they suggested that carbonaceous chondritic impactors delivered this material that produced the large Veneneia basin. The dark material on Vesta's surface is primarily identified by regions of lower albedo and shallower spectral band depth compared to the average Vesta surface. In addition, impact melt on Vesta exhibits moderate albedos, but with suppressed absorption features and redder slopes compared to the average surface (Le Corre et al., 2013).

Compared to Vesta and typical HED/basaltic asteroids, Vestoids that display weaker than normal absorption band depths and lower albedos could provide additional evidence of remnant carbonaceous chondritic material from a primordial impactor or impact melt. Albedo and band depth data from Table 2 and Table 4 show that the three asteroids with suppressed absorption band depths [(3867) Shiretoko, (6331) 1992 FZ1, (30872) 1992 EM17] do not exhibit lower geometric or infrared albedos (Masiero et al., 2011).

Surface roughness on $V_p$-type asteroids may also play a role as Benner et al. (2008) showed that these asteroids display high radar circular polarization ratios (SC/OC) that indicate significant surface roughness on centimeter-to-decimeter particle size scales. Asteroid surfaces with larger average particles sizes may exhibit absorption feature depths that are shallower compared to surfaces with smaller average particle sizes. Hence, the variations in absorption band depths cannot yet be attributed to carbonaceous chondritic or other low-albedo carbonaceous or organic material based on the existing WISE-derived albedos for these three asteroids and current knowledge of V-type asteroid surface properties (Benner et al., 2008).

### *3.3. Orbital analysis*

The dynamical and WISE-derived physical properties for the eight asteroids are shown in Table 5, along with the orbital elements for (4) Vesta. Orbital elements were obtained from the JPL Horizons online ephemeris service at: http://ssd.jpl.nasa.gov/horizons.cgi. WISE physical parameters were obtained from Masiero et al. (2011).

All eight asteroids are within $\Delta a$ = ~0.08 AU, $\Delta i$ = ~2.5°, and $\Delta e$ = ~0.08 of (4) Vesta in dynamical space. While (6331) 1992 FZ1 and (17469) 1991 BT are dynamically classified in the Vesta family (Zappala et al., 1995; Nesvorny et al., 2012; Mothe'-Diniz et al., 2012), four asteroids have no family classification [(3867) Shiretoko, (6976) Kanatsu, (29796) 1999 CW77, (30872) 1992 EM17], and two asteroids are classified in the Flora dynamical family [(5235) Jean-Loup, (5560) Amytis: Zappala et al., 1995; Nesvorny et al., 2012)].

Zappala et al. (1995) plotted proper orbital elements for several asteroid families that include the Vesta and Flora families. In a plot of proper sin *i'* vs. proper *a'*, Zappala et al. (1995) show that the Vesta and Flora families are in close proximity in proper *a'* (~2.3 AU) and in proper sin *i'* (~0.12). This close proximity between the two dynamical asteroid families is the likely cause for (5235) Jean-Loup and (5560) Amytis being classified in the Flora family (Zappala et al., 1995; Nesvorny et al., 2012). Howardite meteorite analog interpretations for these two asteroids suggest that they have migrated into the adjacent Flora family via small body forces such as the Yarkovsky and YORP effects (Bottke et al., 2006). WISE-derived effective diameters for the asteroids are < 8 km, which is within the size range applicable to these non-gravitational forces (Bottke et al., 2006).

Another possible interpretation is that (5235) Jean-Loup and (5560) Amytis are surface crustal fragments from (8) Flora if Flora represents the mantle/core of a fully differentiated parent body that was later disrupted (Gaffey, 1984). (5235) Jean-Loup and (5560) Amytis could then represent fragments from the basaltic crust of the original (8) Flora parent body that had been fully differentiated and later disrupted. Recent spectral and mineralogical work on (8) Flora and the Flora dynamical family, however, is consistent with a non-igneous, LL-chondrite interpretation (Reddy et al., 2014 and references therein). An LL-chondrite interpretation would then suggest that (5235) Jean-Loup and (5560) Amytis are Vesta interlopers into the Flora dynamical family.

| Asteroid | *a* | *e* | *i* | WISE $D_{eff}$ (km) | WISE $p_v$ | WISE $p_{IR}$ |
|---|---|---|---|---|---|---|
| (3867) Shiretoko | 2.351 | 0.107 | 6.273° | 5.345 ± 0.153 | 0.324 ± 0.058 | 0.487 ± 0.087 |
| (5235) Jean-Loup | 2.297 | 0.142 | 4.849° | 7.391 ± 0.024 | 0.301 ± 0.067 | 0.435 ± 0.034 |
| (5560) Amytis | 2.286 | 0.108 | 5.619° | 4.703 ± 0.041 | 0.290 ± 0.051 | 0.451 ± 0.090 |
| 6331 (1992 FZ1) | 2.358 | 0.133 | 7.762° | 5.321 ± 0.101 | 0.473 ± 0.101 | 0.554 ± 0.115 |
| (6976) Kanatsu | 2.332 | 0.169 | 8.247° | 5.497 ± 0.116 | 0.307 ± 0.042 | 0.460 ± 0.064 |
| 17469 (1991 BT) | 2.371 | 0.084 | 6.163° | 5.999 ± 0.173 | 0.258 ± 0.035 | 0.386 ± 0.052 |
| 29796 (1999 CW77) | 2.344 | 0.075 | 7.872° | 4.851 ± 0.073 | 0.248 ± 0.062 | 0.392 ± 0.025 |
| 30872 (1992 EM17) | 2.325 | 0.115 | 5.061° | 3.146 ± 0.080 | 0.409 ± 0.050 | 0.658 ± 0.083 |
| (4) Vesta | 2.362 | 0.090 | 7.134° | -- | -- | -- |

Table 5. Orbital and WISE-derived physical parameters for the eight WISE-defined $V_p$-type asteroids in this paper and orbital elements for (4) Vesta.

## 4. Conclusions and future work

The combination of the above analyses leads to the following conclusions: 1) the eight asteroids in this paper are broadly consistent with a basaltic achondrite/HED meteorite analog and an origin on the surface of (4) Vesta; 2) (3867) Shiretoko exhibits spectral characteristics most similar to the eucrites and may represent an ancient surface or shallow subsurface basaltic flow on (4) Vesta that was later ejected; 3) (5235) Jean-Loup, (5560) Amytis, (6331) 1992 FZ1, (6976) Kanatsu, (17469) 1991 BT, (29796) 1999 CW77, and (30872) 1992 EM17 are spectrally consistent with the howardites (± possible eucrite or diogenite enhancement) that represent surface breccias on Vesta that were ejected via impacts; 4) the WISE-derived albedos for the asteroids are broadly consistent with basaltic surfaces, but with a few albedos [(17469) 1991 BT and (29796) 1999 CW77] that appear somewhat lower than expected; 5) the $V_p$ taxonomic class successfully predicted that 100% of the asteroids in this study (8/8) have basaltic/HED characteristics; 6) the asteroids are all in dynamically favorable locations near (4) Vesta; and 7) all eight of these WISE-defined $V_p$-type asteroids are probable genetic Vestoids.

This is the initial effort in a larger project that seeks to characterize a large sample (~139) of WISE-defined $V_p$-type asteroids. Future work will observe and analyze additional $V_p$-type asteroids that are dynamically near (4) Vesta as well as those $V_p$-type asteroids that are beyond the 3:1 mean-motion resonance in the mid- to outer-regions of the main belt. Results will help to: 1) better constrain the ability of the Carvano et al. (2010) $V_p$-type taxonomy in predicting basaltic/HED-like asteroid surface mineralogies; 2) better constrain the genetic Vestoid population and consequent current Vestoid mass estimates; and 3) better constrain the population of outer main-belt basaltic asteroids and their overall current mass estimate.

## 5. Acknowledgements

The authors thank the anonymous reviewers for their very helpful comments that improved this paper. The authors also gratefully acknowledge the continuing assistance of the NASA Infrared Telescope Facility (IRTF) telescope operators and staff, whose help has been invaluable in obtaining high-quality, publishable asteroid data. We thank Sean Lindsay for the SARA/IDL data reduction program and the time he took to teach us how to use the program. We thank Russ Genet for joining us at the NASA IRTF for the observing run and for his enthusiasm to learn how to observe asteroids as well as to scheme about conducting future research projects. Finally, we thank Denise Meeks for lending her copy editing expertise to this manuscript.

This work is supported by NASA Planetary Astronomy Program grant #NNX14AJ37G. This publication makes use of data products from the Wide-Field Infrared Survey Explorer, which is a joint project of the University of California, Los Angeles, and the Jet Propulsion Laboratory/California Institute of Technology, funded by the National Aeronautics and Space Administration.